\documentstyle[12pt]{article}
\newcommand{\newc}{\newcommand} 
\newc{\ra}{\rightarrow} 
\newc{\lra}{\leftrightarrow} 
\newc{\beq}{\begin{equation}} 
\newc{\eeq}{\end{equation}} 
\newc{\barr}{\begin{eqnarray}} 
\newc{\earr}{\end{eqnarray}} 

 1

\begin{document} 
\begin{titlepage}


\vspace*{.3cm}
\begin{center}
{\large \bf LSP-NUCLEUS ELASTIC SCATTERING CROSS SECTIONS 
\footnote{ Presented by J.D. Vergados}}

\vspace{.5cm}

{J.D. VERGADOS and T.S. KOSMAS }
\end{center}

\vspace{.3cm}
\centerline{\it Theoretical Physics Section, University of Ioannina, 
 GR 451 10, Greece}

\vspace{.4cm}

\abstract{We calculate LSP-nucleus elastic scattering cross sections using 
some representative input in the restricted SUSY parameter space.
The coherent matrix elements are computed throughout the periodic table
while the spin matrix elements for the proposed $^{207}Pb$ target which has 
a rather simple nuclear structure. The results are compared 
to those given from other cold dark matter detection targets. }

\end{titlepage}


\section{Introduction }
There are many arguments supporting the fact that, the cold dark matter 
of the universe, i.e. its component which is composed of particles which 
were non-relativistic at the time of structure formation, is at least
$60\%$.~\cite{Jungm,COBE} There are two interesting cold dark
matter candidates:
i) Massive Compact Halo Objects (MACHO's) and
ii) Weak Interacting Massive Particles (WIMP's).
In the present work we discuss a special WIMP candidate connected with
the supersymmetry, i.e. the lightest supersymmetric particle (LSP).

We examine the possibility to directly detect the LSP via the recoiling
of a nucleus (A,Z) in the elastic scattering process:
$\,\,\chi \, +\, (A,Z) \, \ra \, \chi \,  + \, (A,Z)^* \,\, $
($\chi$ denotes the LSP). In this investigation, we proceed with the 
following steps:

\noindent 
1) We write down the effective Lagrangian at the elementary particle 
(quark) level obtained in the framework of supersymmetry as described 
in refs.~\cite{JDV,KVprd}

\noindent 
2) We go from the quark to the nucleon level using an appropriate quark 
model for the nucleon. Special attention in this step is paid to the 
scalar couplings, which dominate the coherent part of the cross section
and the isoscalar axial currents which depend on the assumed quark model.

\noindent 
3) We compute the relevant nuclear matrix elements using 
as reliable as possible many boby nuclear wave function hopping
that, by putting as accurate nuclear physics input as possible, 
we will be able to constrain the SUSY parameters as much as possible.

\noindent 
4) We calculate the modulation of the cross sections due to the earth's
revolution around the sun and the motion of the sun around the center of
the galactic disc by a folding procedure.

There are many popular targets for LSP detection as $^{19}F$, $^{23}Na$,
$^{29}Si$, $^{40}Ca$, $^{73,74}Ge$, $^{127}I$, $^{207}Pb$ etc.
Among them $^{207}Pb$ has been recently proposed~\cite{KVprd} as an 
important detector, 
since its spin matrix element, especially the isoscalar one, does 
not exhibit large quenching
as that of the light and up to now much studied $^{29}Si$ and $^{73}Ge$ 
nuclei.~\cite{Ress}

Our purpose is to calculate LSP-nucleus scattering cross section using 
some representative input in the restricted SUSY parameter space, 
to compute the coherent LSP-nucleus scattering cross sections throughout 
the periodic table and study the spin matrix elements of $^{207}Pb$, 
since this target, in addition to its 
experimental qualifications, has the advantage of a rather simple nuclear
structure.  We compare our results to those given from other proposed
cold dark matter detection targets~\cite{Ress} and present results calculated
by using new imput SUSY parameters~\cite{Casta} obtained in a phenomenologically
allowed parameter space.

\section{ Momentum transfer dependence of the total cross sections}

The total cross section of the LSP-nucleus elastic scattering can be written 
as~\cite{KVprd}

\barr
\sigma &=& \sigma_0 (\frac{m_1}{m_p})^2 \frac{1}{(1+\eta)^2} \,
 \{ A^2 \, [[\beta^2 (f^0_V - f^1_V \frac{A-2 Z}{A})^2 
\nonumber \\ & + & 
(f^0_S - f^1_S \frac{A-2 Z}{A})^2 \, ]I_0(u) -
\frac{\beta^2}{2} \frac{2\eta +1}{(1+\eta)^2}
(f^0_V - f^1_V \frac{A-2 Z}{A})^2 I_1 (u) ]
\nonumber \\ & + & 
(f^0_A \Omega_0(0))^2 I_{00}(u) + 2f^0_A f^1_A \Omega_0(0) \Omega_1(0)
I_{01}(u) + (f^1_A \Omega_1(0))^2 I_{11}(u) \, \} 
\label{eq:2.1}
 \earr

\noindent 
where $m_1, \, m_p\,$ is the LSP, proton masses, $\eta = m_1/m_p A$ and
$\sigma_0 \simeq 0.77 \times 10^{-38}cm^2$. The parameters $f^{\tau}_S, 
\,\,f^{\tau}_V, \,\, f^{\tau}_A$, with $\tau =0,1$ an isospin index, 
describe the scalar, vector and axial vector couplings and 
are determined by a specific SUSY mechanism.~\cite{Kane,Casta}
The momentum transfer enters via u as

\beq
u = \frac{1}{2} \left( \frac{2\beta m_1 c^2}{(1+\eta)} 
\frac{b}{\hbar c}\right)^2,
\qquad \beta = \frac{v}{c} \, \approx \, 10^{-3}
\label{eq:2.2}  
\eeq

\noindent 
(b is the harmonic oscillator parameter).
In eq. (\ref{eq:2.1}), the terms in the square brackets describe the coherent
cross section and involve the integrals $I_{\rho}(u)$ ($\rho =0,1$)
while the other terms describe the spin dependence of the  
cross section by means of the integrals  $I_{\rho\rho^{\prime}}(u)$
($\rho,\rho^{\prime} =0,1$, isospin indices) which are normalized so 
as $I_{\rho}(0)=1$ and $I_{\rho \rho^{\prime}}(0)=1$. 
For their definitions see ref.~\cite{KVprd} where they are calculated 
by using realistic nuclear form factors.

In ref.~\cite{KVprd} we studied the variation of $I_{\rho}$ as functions 
of the LSP mass. We found that $u$ can be quite big for large mass of the 
LSP and heavy nuclei even though the energy transfer is small ($\le 100 
KeV$). The total cross section can in such instances be reduced by a factor 
of about five. 

The spin matrix element of heavy nuclei like $^{207}Pb$ 
has not been evaluated, since one expects the relative 
importance of the spin versus the coherent mode to be more pronounced
on light nuclei. However, the spin matrix element in the 
light isotopes is quenched, while that 
of $^{207}Pb$ does not show large quenching. For this feature,
we recently proposed $^{207}Pb$ nucleus as an important candidate
in the LSP detection.~\cite{KVprd} Furthermore, this nucleus has some 
additional advantages
as: i) it is believed to have simple structure (one $2p 1/2$ neutron
hole outside the doubly magic nucleus $^{208}Pb$) and 
ii) it has low angular momentum and therefore only two multipoles $\lambda =0$
and $\lambda =2$ with a $J$-rank of $\kappa=1$ can contribute even at large
momentum transfers. 

The dependence of the total cross section on the momentum transfer for 
$^{207}Pb$ has been studied in ref.~\cite{KVprd} where the variation of 
the integrals $I_{\rho\rho^{\prime}}(u)$, which describe the spin part of 
$^{207}Pb$, has also been investigated.

\section{Folding of the cross section with a Maxwellian distribution}
Due to the revolution of the earth around the sun, the LSP-nucleus 
cross sections appear modified. This effect is studied by folding
the total cross section eq. (\ref{eq:2.1}) with an appropriate distribution. 
If we assume a Maxwell type distribution, which is consistent with
the velocity distribution of LSP into the galactic halo,
the counting rate for a target with mass $m$ takes the form~\cite{KVprd}

\beq
\Big<\frac{dN}{dt}\Big> =\frac{\rho (0)}{m_1} \frac{m}{Am_p} \sqrt{<v^2>}
<\Sigma>
\label{eq:3.1}  
\eeq

\noindent
with $\rho (0) = 0.3 GeV/cm^3$, the LSP density in our vicinity, and
$<\Sigma>$ given by

\barr
<\Sigma>&=&\Big(\frac{m_1}{m_p}\Big)^2 \frac{\sigma_0}{(1+\eta)^2}
\Big\{A^2 \Big[ <\beta^2> \\
\nonumber
&& \times \Big(f^0_V-f^1_V \frac{A-2 Z}{A})^2 
\Big(J_0-\frac{2\eta+1}{2(1+\eta)^2}J_1\Big) +
(f^0_S-f^1_S\frac{A-2 Z}{A})^2{\tilde J}_0\Big]  \\
\nonumber
&& + \Big( f^0_A \Omega_0(0)\Big)^2 J_{00}
+ 2 f^0_A f^1_A \Omega_0(0)\Omega_1(0) J_{01}
+ \Big( f^1_A \Omega_1(0)\Big)^2 J_{11} \Big\}
\label{eq:3.2}  
\earr

\noindent
The parameters ${\tilde J}_0$, $J_\rho$, $J_{\rho\sigma}$ describe the
scalar, vector and spin part of the velocity averaged counting rate, 
respectively. They are functions of $\beta_0 =v_0/c$, $\delta =2v_1/v_0$
and $u_0$, where $v_0$ is the velocity of the sun around the galaxy,
$v_1$ the velocity of the Earth around the sun and $u_0$ is given
by an expression like that of eq. (\ref{eq:2.2}) with $\beta \rightarrow 
\beta_0$. Since $\delta \approx 0.27 << 1 $, we can expand the J-integrals
in powers of $\delta$ and retain terms up to linear in $\delta$. Thus,
for each mechanism (vector, scalar, spin) we obtain two integrals 
associated with $l=0$ and $l=1.$ The most important $K^l$ integrals are 
studied in ref.~\cite{KVprd} By exploiting the above expansion of K-integrals,
the counting rate can be written in the form

\beq
\Big<\frac{dN}{dt}\Big>=\Big<\frac{dN}{dt}\Big>_{0}(1 + h \, cos\alpha)
\label{eq:3.3}
\eeq

\noindent
where $\alpha$ is the phase of the earth's orbital motion and 
$\big<dN/dt\big>_0$ is the rate obtained from the $l=0$ multipole. 
$h$ represent the amplitude of the oscillation, i.e. the ratio of the component
of the multipole $l=1$ to that of $l=0$. 

\section{Results and discussion }
The three basic ingredients of our calculation were 
the input SUSY parameters, a quark model for the nucleon and the structure
of the nuclei involved.
The input SUSY parameters used for the results presented in Tables 1 and 2
have been calculated in a phenomenologically allowed 
parameter space (cases \#1, \#2, \#3) as explained in ref.~\cite{KVprd}

\begin{table}[t]
\caption{Comparison of the static spin matrix elements for three
typical nuclei.\label{tab:1}}
\vspace{0.4cm}
\begin{center}
\footnotesize
\begin{tabular}{|l|rrr|}
\hline
Component & $^{207}Pb_{{1/2}^-}$ & $^{73}Ge_{{9/2}^+}$ 
 & $^{29}Si_{{1/2}^+}$ \\
\hline
$\Omega^2_1(0)$ & 0.231 & 1.005 & 0.204 \\
$\Omega_1(0) \Omega_0(0)$ & -0.266 & -1.078 & -0.202 \\
$\Omega^2_0(0)$ & 0.305 & 1.157 & 0.201 \\
\hline
\end{tabular}
\end{center}
\end{table}

For the coherent part (scalar and vector) we used realistic nuclear
form factors and studied three nuclei, representaves
of the light, medium and heavy nuclear isotopes ($Ca$, $Ge$ and $Pb$).
In Table 1 we show the results obtained for three different quark
models denoted by A (only quarks u and d) and B, C (heavy quarks
in the nucleon).
We see that the results vary substantially and are sensitive to 
the presence of quarks other than u and d into the nucleon.

\begin{table}[t]
\caption{The quantity $<dN/dt>_0$ in $y^{-1}Kg^{-1}$ and the 
oscillator parameter h for the coherent vector and scalar 
contributions. 
\label{tab:2}}
\vspace{0.4cm}
\begin{center}
\footnotesize
\begin{tabular}{|rl|cc|lrrc|}
\hline
 & & \multicolumn{2}{|c|}{Vector $\,\,\,$ Contribution}  &
     \multicolumn{4}{|c|}{Scalar $\,\,\,$ Contribution}  \\
\hline 
& & $<dN/dt>_0$ &$h$ & \multicolumn{3}{c}{$<dN/dt>_0 $} & $h$ \\ 
\hline
& \multicolumn{1}{c|}{Case} &$(\times 10^{-3})$ &  &
    \multicolumn{1}{c}{ Model $\,\,$ A }& 
    \multicolumn{1}{c}{ Model $\,\,$ B }& 
    \multicolumn{1}{c}{ Model $\,\,$ C }&  \\ 
\hline
  &$\#1$& 0.264 &0.029 &$0.151\times 10^{-3}$ &  0.220 &  0.450 &-0.002 \\
Pb&$\#2$& 0.162 &0.039 &$0.410\times 10^{-1}$ &142.860 &128.660 & 0.026 \\
  &$\#3$& 0.895 &0.038 &$0.200\times 10^{-3}$ &  0.377 &  0.602 &-0.001 \\
\hline
  &$\#1$& 0.151 &0.043 &$0.779\times 10^{-4}$ &  0.120 &  0.245 & 0.017 \\
Ge&$\#2$& 0.053 &0.057 &$0.146\times 10^{-1}$ & 51.724 & 46.580 & 0.041 \\
  &$\#3$& 0.481 &0.045 &$0.101\times 10^{-3}$ &  0.198 &  0.316 & 0.020 \\
\hline
  &$\#1$& 0.079 &0.053 &$0.340\times 10^{-4}$ &  0.055 &  0.114 & 0.037 \\
Ca&$\#2$& 0.264 &0.060 &$0.612\times 10^{-2}$ & 22.271 & 20.056 & 0.048 \\
  &$\#3$& 0.241 &0.053 &$0.435\times 10^{-4}$ &  0.090 &  0.144 & 0.038 \\
\hline
\end{tabular}
\end{center}
\end{table}

\begin{table}[t]
\caption{The spin contribution in the $LSP-^{207}Pb$ scattering 
for two cases: EMC data and NQM Model. The LSP mass is
$m_1=126, \,\,27, \,\,102 \,\,GeV$ for $\#1, \#2, \#3$ 
respectively. \label{tab:3}}
\vspace{0.4cm}
\begin{center}
\footnotesize
\begin{tabular}{|l|ll|lc|}
\hline
& \multicolumn{2}{|c}{\hspace{1.2cm}EMC \hspace{.2cm} DATA} \hspace{.8cm} &
 \multicolumn{2}{c|}{\hspace{1.2cm}NQM \hspace{.2cm} MODEL} \\ 
\hline
Solution & \hspace{.2cm}$< dN/dt >_0 $ $(y^{-1} Kg^{-1})$
  \hspace{.2cm} & $ h $ & 
$\hspace{.2cm} < dN/dt >_0 $ $ (y^{-1} Kg^{-1})$ & $ h $ \\ 
\hline
$\#1  $  &$0.285\times 10^{-2}$& 0.014 &$0.137\times 10^{-2}$& 0.015  \\
$\#2  $  & 0.041               & 0.046 &$0.384\times 10^{-2}$& 0.056  \\
$\#3  $  & 0.012               & 0.016 &$0.764\times 10^{-2}$& 0.017  \\
\hline
\end{tabular}
\end{center}
\end{table}

The spin contribution, arising from the axial current,
was computed in the case of $^{207}Pb$ system. 
For the isovector axial coupling the transition from the quark to
the nucleon level is trivial (a factor of $g_A=1.25)$. For the
isoscalar axial current we considered two possibilities 
depending on the portion of the nucleon spin which is attributed to the
quarks, indicated by EMC and NQM.~\cite{KVprd}
The ground state wave function of $^{208}Pb$
was obtained by diagonalizing the nuclear 
Hamiltonian in a 2h-1p space which is standard for this doubly magic nucleus.   
The momentum dependence of the matrix elements was taken into account and all 
relevant multipoles were retained (here only $\lambda=0$ and $\lambda=2$).

In  Table 3, we compare the spin matrix elements at $q=0$ for the 
most popular targets considered for LSP detection $^{207}Pb$, 
$^{73}Ge$ and $^{29}Si$. We see that, even though the spin matrix elements 
$\Omega^2$ are even a factor of three smaller than those for $^{73}Ge$ obtained 
in ref.~\cite{Ress} (see Table 1), their contribution to the total
cross section is almost the same (see Table 4) for LSP masses around 
$100 \, GeV$. Our final results for the quark models (A, B, C, NQM, EMC)
are presented in Tables 3 and 4 for SUSY models \#1-\#3~\cite{Kane}
and Tables 4 and 5 for SUSY models \#4-\#9.~\cite{Casta}

\begin{table}[t]
\caption{Ratio of spin contribution ($^{207}Pb/^{73}Ge$) at the relevant
momentum transfer with the kinematical factor $1/(1+\eta)^2, \,\, 
\eta=m_1/A m_p.$
\label{tab:4}}
\vspace{0.4cm}
\begin{center}
\footnotesize
\begin{tabular}{|c|lllllllll|}
\hline
 Solution & $\#$1 & $\#$2 & $\#$3 & $\#$4 & $\#$5 & $\#$6 & $\#$7 & $\#$8 
 & $\#$9 \\
\hline
$m_1 \,(GeV)$ & 126 & 27 & 102 & 80 & 124 & 58 & 34 & 35 & 50 \\
\hline
NQM & 0.834 & 0.335 & 0.589 & 0.394 & 0.537 & 0.365 & 0.346 & 0.337 & 0.417 \\
EMC & 0.645 & 0.345 & 0.602 & 0.499 & 0.602 & 0.263 & 0.341 & 0.383 & 0.479 \\
\hline
\end{tabular}
\end{center}
\end{table}

\begin{table}[t]
\caption{The same parameters as in Tables 1 and 2. The LSP mass
is $m_1= 80,$ 124, 58, 34, 35, 50 $GeV$ for the cases $\#4-\#9$ 
respectively. Cases $\#8, \#9$ are no-scale models. The values of 
$<dN/dt>_0$ for Model A and the Vector part must be multiplied by
$\times 10^{-2}$.
\label{tab:5}}
\vspace{0.4cm}
\begin{center}
\footnotesize
\begin{tabular}{|l|crrr|lr|lll|}
\hline
  & \multicolumn{4}{|c|}{Scalar$\,$ Part} & 
    \multicolumn{2}{c|}{Vector $\,$ Part} &
    \multicolumn{3}{c|}{Spin   $\,$ Part} \\
\hline 
& \multicolumn{3}{c}{$\big<\frac{dN}{dt}\big>_0$}& $h$ & 
$\big<\frac{dN}{dt}\big>_0$ & $h$ & \multicolumn{2}{c}
{$\big<\frac{dN}{dt}\big>_0$} & $h$ \\ 
\hline
& A & B & C & & & & EMC & NQM & \\ 
\hline
$\#4 $& 0.03 & 22.9& 8.5 & 0.003& 0.04 & 0.054
 & $0.80\, 10^{-3}$& $0.16\, 10^{-2}$& 0.015 \\
$\#5 $& 0.46 & 1.8& 1.4 & -0.003& 0.03 & 0.053
 & $0.37\, 10^{-3}$& $0.91\, 10^{-3}$& 0.014 \\
$\#6 $& 0.16 & 5.7& 4.8 & 0.007& 0.11 & 0.057
 & $0.44\, 10^{-3}$ & $0.11\, 10^{-2}$& 0.033 \\
$\#7 $&  4.30 & 110.0& 135.0 & 0.020& 0.94 & 0.065
 & 0.67 & 0.87 & 0.055 \\
$\#8 $& 2.90 & 73.1& 79.8 & 0.020& 0.40 & 0.065
 & 0.22 & 0.35 & 0.055 \\
$\#9 $&  2.90 & 1.6& 1.7 & 0.009& 0.95 & 0.059
 & 0.29 & 0.37 & 0.035 \\
\hline
\end{tabular}
\end{center}
\end{table}

\section{Conclusions }
In the present study we found that for heavy LSP and heavy nuclei
the results are sensitive to the momentum transfer as well as to the
LSP mass and other SUSY parameters. From the Tables 3 and 4 we see that,
the results are also sensitive to the quark structure of the nucleon
in the following sense: i) The coherent scalar 
(associated with Higgs exchange) for model A (u and
d quarks only) is comparable to the vector coherent contribution.
Both are at present undetectable. ii) For models B and C (heavy
quarks in the nucleon) the coherent scalar contribution is
dominant. Detectable rates $<dN/dt>_0 \ge 100\, \, y^{-1}Kg^{-1}$ 
are possible in a number of models with light LSP.

The spin contribution is sensitive to the nuclear structure.
It is undetectable if the LSP is primarily a gaugino. 
The folding of the event rate with the velocity distribution provides 
the modulation effect $h$. In all cases it is small, less than $\pm 5\%$.



\begin{thebibliography}{99} 
\bibitem{Jungm}G. Jungman {\it et al.}, Phys. Rep., {\bf 267} (1996) 195.
\bibitem{COBE}G.F. Smoot et al., (COBE data,) Astrophys. J. {\bf 396}
(1992) L1.
\bibitem{JDV}J.D. Vergados, J. of Phys. {\bf G 22} (1996) 253.
\bibitem{KVprd}T.S. Kosmas and J.D. Vergados, Phys. Rev. {\bf D}, 
to be published.
\bibitem{Smith}P.F. Smith and J.D. Lewin, Phys. Rep. {\bf 187} (1990) 203. 
\bibitem{Kane}G.L. Kane {\it et al.}, Phys. Rev. {\bf D 49} (1994) 6173.
\bibitem{Ress}M.T. Ressell 
{\it et al.}, 
Phys. Rev. {\bf D 48} (1993) 5519.
\bibitem{Casta}D.J. Casta\v no, E.J. Piard and P. Ramond, Phys. Rev. 
{\bf D 49} (1994) 4882;

D.J. Casta\v no, Private Communication. 
\end{thebibliography}
\end{document}